\documentstyle[aps,preprint,eqsecnum,epsfig,rotate]{revtex}
\newcommand{\One}{1\kern-4.5pt1}
\tightenlines
\begin{document}
\draft
\title{ Symmetries and Spectrum of 
SU(2) Lattice Gauge Theory 
at Finite Chemical  Potential} 
\author{
Simon Hands$^a$, John B. Kogut$^b$, Maria-Paola Lombardo$^c$, 
Susan E. Morrison$^a$}

\address{$^a$ Department of Physics, University of Wales Swansea,\\
Singleton Park, Swansea, SA2 8PP, U.K.}
\address{$^b$ Department of Physics,
University of Illinois at Urbana-Champaign\\
1110 West Green Street,
Urbana, IL 61801-3080, U.S.A.}
\address{$^c$ Istituto Nazionale di Fisica Nucleare, 
Laboratori Nazionali del Gran Sasso, \\I-67010 Assergi (AQ), Italy
\footnote{Current address} \\
and\\
Fakult\"at f\"ur Physik, Universit\"at Bielefeld, Postfach 100 31, 
D-33501 Bielefeld, Germany}

\maketitle

\begin{abstract}

We study SU(2) Lattice Gauge Theory with dynamical 
fermions at non-zero chemical potential $\mu$.
The symmetries special to SU(2) for staggered fermions on the lattice are 
discussed
explicitly and their relevance to spectroscopy and condensates at non-zero 
chemical potential are considered. Using the molecular dynamics algorithm on 
small lattices we
find qualitative changes in the theory's spectroscopy 
 at small and
large values of $\mu$.
This preliminary
study should lay the groundwork for future large scale simulations.

\end{abstract}

\section{Introduction}

Simulations of gauge theories with dynamical fermions
using the SU(2) colour group at nonzero density 
where the action is still real can be performed using standard
algorithms, such as hybrid molecular dynamics \cite{DuKo}
or hybrid Monte Carlo \cite{DKPR}. 
These algorithms fail dramatically for QCD which has 
a complex action once the quark chemical potential $\mu\not=0$
\cite{KL}. 
Attempts to simulate field theories with complex actions, such as
QCD with a nonzero chemical potential, include 1. Langevin
algorithms, 2. the Glasgow algorithm pioneered by I. Barbour \cite{BarBel92},
and 3. analytic extensions of simulations at imaginary chemical potential.
None of these methods has succeeded. The complex Langevin algorithms
suffer from uncontrollable instabilities, the Glasgow algorithm requires
statistics that grow at least as fast as the exponential 
of the size of the system, and simulations at imaginary chemical 
potential provide no discernable hints of
transitions at real values of $\mu$. These failures are particularly
painful because, unlike other phenomena which are not qualitatively sensitive 
to the gauge group, the physics of field theories at $\mu\not=0$ is crucially
dependent on the gauge group. These points will be made clear 
below where SU(2) will
be considered in detail. Very briefly, baryons and mesons are essentially
identical in the SU(2) theory, so all the special features of QCD at nonzero
chemical potential which depend on qualitatively different meson and baryon
spectroscopy are lost. Of even more significance to the 
scenarios to be discussed below is the fact that diquarks in the SU(2) 
theory can be gauge singlets while they
cannot be singlets in SU(3). So, if we search for and find strong 
diquark correlations and/or condensates in the SU(2) theory, these results 
may have no bearing on such phenomena in the SU(3) theory. The energetics of
condensate formation is SU(3) is more challenging, both conceptually
and computationally, than in SU(2).

All of these points have conspired to lead researchers, both using lattice
methods or  crude analytic methods such as instanton models or
mean field approaches based on gap equations of simplified field theories, to
reconsider SU(2). This will be done below, based on the hope that a controlled 
and
detailed understanding of the SU(2) theory at $\mu\not=0$ will give 
some insight 
into a subset of the phenomena expected in QCD. A quantitative study of QCD at
nonzero $\mu$ must await further breakthroughs.

The simplest scenario for QCD with $\mu\not=0$, which we shall
refer to as the {\sl standard scenario\/},
predicts that as $\mu$ is increased, observables such as
the chiral condensate $\langle\bar q q\rangle$, which is 
non-zero in the vacuum, and quark number density 
$n$ remain constant until
$\mu_{o}\sim m_{lb}/N_c$, 
where $m_{lb}$ denotes the mass of the lightest baryon in
the spectrum,  whereupon a non-zero density of
quarks is induced into the ground state, corresponding to a state of nuclear
(or hadronic) matter.
At some $\mu_c\geq\mu_o$ chiral symmetry should be restored in a ({\sl
first order\/}) transition, resulting in a state of quark matter. 
Above $\mu_c$ there are competing
effects governing the quarks' influence 
on the gluon degrees of freedom. On the one hand, we expect
enhanced screening 
due to quasi-massless virtual $q\bar q$ pairs 
as the chiral symmetries are restored; on the
other, we expect the contribution to screening from 
$q\bar q$ pairs lying deep within the Fermi surface to be suppressed
for kinematic reasons.

\begin{figure}
\begin{center}
{\epsfig{file=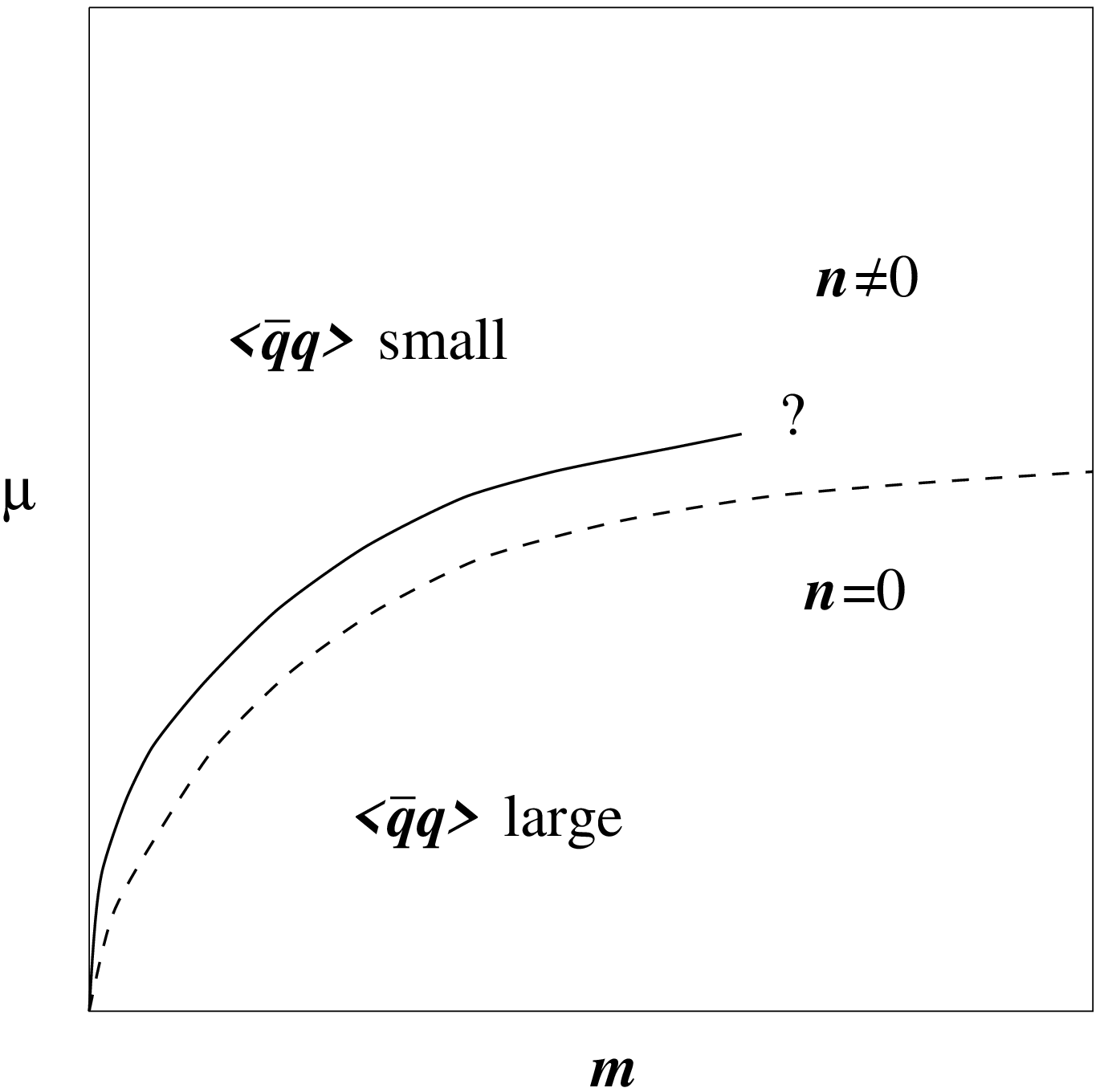, width=12 truecm}}
\end{center}
\caption[xxx]{Tentative phase diagram in the $(m,\mu)$ plane
for the standard scenario 
} \label{fig:phase1}
\end{figure}

In Figure \ref{fig:phase1} we present a 
tentative phase diagram for the standard scenario at zero
temperature in SU(2) in the $(m,\mu)$ plane.
The dotted line corresponds to $\mu_o(m)$, separating a vacuum phase
in which $n\equiv0$ from a phase of ``nuclear matter'' with $n>0$. 
The curvature
of this line reflects the expected behaviour $m_{lb}\propto\sqrt{m}$
(note that the lightest baryon in SU(2) gauge theory is degenerate with 
the pion, and hence becomes massless in the chiral limit by standard PCAC 
arguments).
The solid
line corresponds to $\mu_c(m)$, which again is expected to scale as $\sqrt{m}$:
since the chiral condensate $\langle\bar qq\rangle$ is never vanishing for
$m>0$, it is not clear {\sl a priori\/} whether this line demarcates a genuine
phase separation, or whether there may in principle be an end-point; this
uncertainty is reflected by the question mark (the issue of whether
there is a phase separation between ``confined'' and ``deconfined''
phases is problematic with dynamical fermions). 
Note that in the chiral limit
both critical values of the chemical potential tend to zero.

Further motivation for investigation of the model comes
from  recent speculation that the ground state of dense baryonic
matter may be more exotic \cite{ARW}\cite{RSSV}. On the assumption that
at high density and zero temperature the quarks have a large Fermi surface, 
then it is natural to suppose that Bose condensation of quark pairs in the
vicinity of the surface takes place, leading to the breaking of 
the global symmetry of baryon number, and hence superfluidity, and/or even the 
local gauge symmetry, leading to ``color superconductivity''. This phenomoneon
would be signalled by a non-vanishing {\sl diquark condensate\/} $\langle
qq\rangle$.  
As pointed out above, 
for $N_c=2$ the situation is qualitatively simpler than what might happen in 
$SU(3)$:
since a SU(2) diquark condensate does not break gauge invariance,  its
formation would not lead to 
color superconductivity.

The first lattice analysis of two colors QCD with non-vanishing chemical
potential which pointed out the relevance of diquark condensates was a
quenched simulation \cite{KarDag} performed in the strong coupling
limit $g^2 = \infty$; the results were then compared with 
analytical predictions obtained from a $1/d$ expansion combined with a mean 
field analysis, and further confirmed by (nearly exact)
monomer-dimer simulations \cite{DMW}.  For the SU(2) theory
at $g^2 =\infty$, $T=0$, $\mu \not= 0$ it was concluded that there is {\bf no 
chiral symmetry restoring
transition}. 
The simulations suggested that the chiral condensate $\langle 
\bar qq \rangle$
actually vanishes for all $\mu \not= 0$ in the limit $m \rightarrow 0$,
while $\langle qq\rangle\not=0$ for $\mu>m_{\pi}/2$, thus
spontaneously 
breaking both the $\mbox{U(1)}_V$ 
global symmetry associated with conservation of baryon
number and the chiral symmetry. 
The simulation results were explained
by noticing that the chiral condensate rotates to a
baryonic diquark condensate.
A baryonic source term was introduced in a mean  field 
analysis and in the dimer calculations but not in exact (hybrid) 
simulations. 
The spontaneous breakdown of the $\mbox{U(1)}_V$ 
symmetry is believed to be a direct consequence of the introduction of a 
chemical potential.  A quenched study of the chiral
condensate for two-colors QCD has been published \cite{quensu2}. 

A detailed  analytic study  \cite{CD} confirms  that 
 in the instanton vacuum, at T=0, $\mu > m_\pi/2$ 
the stable phase of the two  colour model is indeed a phase with
diquark condensation, in agreement with the early lattice studies
and the more qualitative discussions  reviewed above.

\begin{figure}
\begin{center}
{\epsfig{file=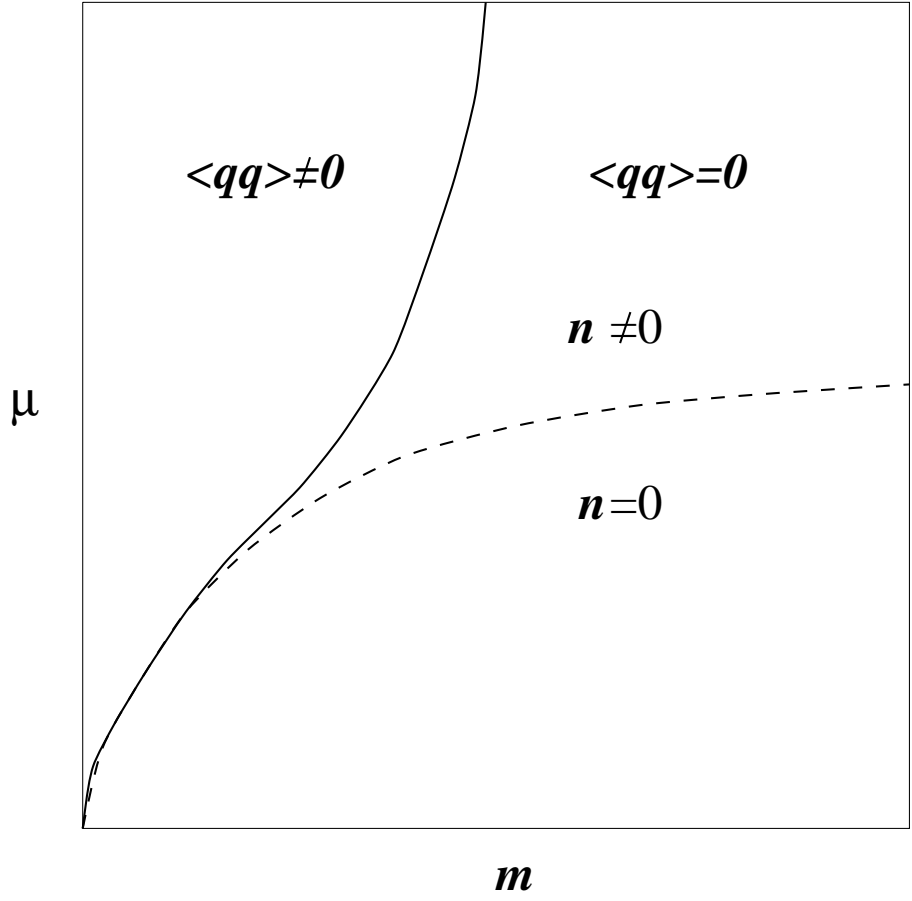, width=12 truecm}}
\end{center}
\caption[xxx]{Tentative phase diagram in the $(m,\mu)$ plane
when a diquark condensate $\langle qq\rangle$ is considered
} \label{fig:phase2}
\end{figure}

In section \ref{sec:symmetries} we shall give a detailed outline of the 
global symmetries of staggered lattice fermions in the fundamental
representation of SU(2), and show that the gauge interaction between
quark pairs is identical to that between quark and anti-quark. 
In the chiral and zero-density limits
a rotation between ground states with $\langle\bar qq\rangle\not=0$ and 
$\langle qq\rangle\not=0$ can thus 
be performed at zero energy cost, and the 
two condensates are physically indistinguishable. For $m\not=0$, $\mu\not=0$, 
the analysis must be refined; now the diquark 
condensate acts as an order parameter signalling the spontaneous breakdown
of baryon number conservation. On the assumption that a diquark condensate
forms at sufficiently high density we find it plausible that a genuine phase
separation occurs, between a normal low density 
phase with $\langle qq\rangle=0$ and
a high density superfluid phase with $\langle qq\rangle\not=0$ (as we shall see
below, the $\langle qq\rangle$ condensate which seems most likely to form 
is gauge 
invariant, and does not give rise to superconductivity in this case). 
This is shown 
as a solid line in the tentative phase diagram of Figure \ref{fig:phase2}; the
curvature of the line this time reflects the fact that at large Fermi energies
asymptotic freedom should weaken the interaction between quark pairs
and thus favour the chiral condensate as $m$ is increased.
Once
again the onset for thermodynamics is expected
to occur for $\mu_o\propto m_\pi$, and this is shown by the dotted line.

By further increasing $\mu > \mu_o \simeq m_\pi/2$  
diquarks could begin to populate the vacuum
producing the ordinary cold
nuclear matter phase of SU(2) gauge theory.  If the dynamics favor pairing,
they could condense. Of course, this phenomenon
would have to compete with the free quark phase. Computer simulations
which are sensitive to relatively small energy differences could, 
in principle, find the separation between 
$\mu_o \simeq m_\pi/2$ and $\mu_c$ and
assess the physical phase between the solid and dotted line
in the phase diagram outlined in Figure \ref{fig:phase2}.  

In this paper we present an analysis of the symmetries of 
the staggered lattice fermion action. We identify Goldstone modes
appropriate to the staggered fermion lattice action in the regimes of low
and high baryon density, which enables a tentative classification
of the spectrum of light excitations in each regime. 
We then present numerical results for 
unquenched  simulations of two colours QCD at $\beta=1.3$ on small lattices, 
which suggests that for $N_c=2$ the
rotation of the chiral condensate into a diquark condensate observed
at infinite coupling is likely to occur also at this coupling, and we
establish the level ordering in the particle spectrum in the high energy 
phase. 

\section{Action and Symmetries}
\label{sec:symmetries}

We will outline plausible patterns of
symmetry breaking for SU(2) lattice gauge theory with 
fermions in the fundamental 
representation of the gauge group. 
We emphasize the distinction  between the model 
with continuum fermions in a pseudoreal representation 
(such as the {\bf2} of SU(2)) which has the symmetry breaking pattern
$\mbox{SU}(2N_f)\to\mbox{Sp}(2N_f)$ in contrast to the lattice model
with staggered fermions which, as we shall outline below, has the 
unorthodox pattern $\mbox{SU}(2N_f)\to\mbox{O}(2N_f)$. We also consider the 
consequences of introducing a chemical potential corresponding to
non-zero baryon density.

Let us start with the kinetic term in the lattice action
for a gauged isospinor doublet ({\bf2})
of staggered fermions $\chi,\bar\chi$. 
For clarity we will consider $N=1$ flavors,
although this will be generalised when the numerical simulations are discussed.
\begin{eqnarray}
S_{kin}=& &\sum_{x,\,\nu=1,3}{{\eta_\nu(x)}\over2}\left[
\bar\chi(x)U_\nu(x)\chi(x+\hat\nu)-\bar\chi(x)U_\nu^\dagger(x-\hat\nu)
\chi(x-\hat\nu)\right]  \nonumber \\
&+ &\sum_{x}{{\eta_t(x)}\over2}\left[
\bar\chi(x)e^{\mu}U_{t}(x)\chi(x+\hat{t})-\bar\chi(x)e^{-\mu}U_{t}^\dagger
(x-\hat{t})
\chi(x-\hat{t})\right]
\label{eq:sfund}
\end{eqnarray}
The $U$'s are $2\times2$ complex matrices acting on color indices.
The chemical potential enters via the timelike links in the standard
way. It is straightforward to rearrange (\ref{eq:sfund}),
using the Grassmann nature of $\chi,\bar\chi$,
the fact that $\eta_\mu(x)
\equiv(-1)^{x_0+\cdots=x_{\mu-1}}=\eta_\mu(x\pm\hat\mu)$, 
and the property
$\tau_2U_\mu\tau_2=U_\mu^*$, where $\tau_2$ is a Pauli matrix:
\begin{eqnarray}
S_{kin}&=&\sum_{x\;even,\,\nu=1,3}{{\eta_\nu(x)}\over2}\left[
\bar X_e(x)U_\nu(x)X_o(x+\hat\nu)-\bar X_e(x)U_\nu^\dagger(x-\hat\nu)
X_o(x-\hat\nu)\right] \nonumber \\
&+ &\sum_{x\;even}{{\eta_{t}(x)}\over2}\left[
\bar X_e(x)\left(\matrix{e^{\mu}&0\cr0&e^{-\mu}}\right) U_{t}(x)X_o(x+\hat{t})
- - - \right. \nonumber \\
& &\;\;\;\;\;\;\;\;\;\;\;\;\;\;\;\;\;\;\left.\bar 
X_e(x)\left(\matrix{e^{-\mu}&0\cr0&e^{\mu}}\right)U_{t}^\dagger(x-\hat{t})
X_o(x-\hat{t})\right]
\label{eq:sfundX}
\end{eqnarray}
with the definitions
\begin{equation}
\bar X_e=(\bar\chi_e,-\chi_e^{tr}\tau_2)\;\;:\;\;
X_o=\left(\matrix{\chi_o\cr -\tau_2\bar\chi_o^{tr}\cr}\right).
\label{eq:Xfields}
\end{equation}
The sum in (\ref{eq:sfundX}) is only over {\sl even\/} sites,
ie. those defined by $\varepsilon(x)=(-1)^{x_1+x_2+x_3+x_4}=+1$,
and the subscripts emphasise that the $\bar X$ fields are only defined
on even sites, and the $X$ on odd.

Now, in the original action (\ref{eq:sfund}) there are two manifest global
U(1) symmetries:
\begin{equation}
\chi\mapsto e^{i\alpha}\chi\;\;\bar\chi\mapsto\bar\chi 
e^{-i\alpha}\;\;\;:\;\;\;
\chi\mapsto e^{i\alpha\varepsilon}\chi\;\;
\bar\chi\mapsto\bar\chi e^{i\alpha\varepsilon},
\end{equation}
the first ($\mbox{U(1)}_V$) corresponding to 
conservation of baryon number and the second, which holds only in the chiral
limit $m=0$ and 
which we will call $\mbox{U(1)}_A$, corresponds to a form of
axial charge conservation. 
In (\ref{eq:sfundX}) we can see that in the chiral limit at zero density
(ie. for $\mu=m=0$) both 
are subsumed in a larger U(2) symmetry
(because exact global symmetries are non-anomalous in lattice formulations,
the full symmetry is U(2); in a 
continuum approach the axial anomaly would reduce it to SU(2)):
\begin{equation}
X_o\mapsto VX_o\;\;\bar X_e\mapsto\bar X_eV^\dagger\;\;\;\;V\in\mbox{U(2)}.
\label{eq:U2}
\end{equation}
By introducing
a non-zero chemical potential we reduce the U(2) symmetry of the 
lattice action to $\mbox{U(1)}_V\otimes\mbox{U(1)}_A$, and by introducing
a bare mass we reduce it to $\mbox{U(1)}_V$. 
This will be important
when we identify and compare the Goldstone modes present in the low and high 
density phases. 

\subsection{Qualitative Discussion of Symmetries and Spectrum}

\subsubsection{$\mu = 0.0$}

As anticipated in the Introduction,  
when $\mu=0$ quarks and antiquarks belong to equivalent representations:
there is a symmetry transformation which turns
a chiral condensate into a diquark one. In other words,  there
is only one discernable condensate, $<\bar q q>^2 + <q q >^2$,
whose orientation in the chiral sphere is selected by the explicit
breaking term.

Recall that the extra quark-antiquark symmetry enlarges the continuous 
global symmetry 
of the lattice Action from $U(f) \times U(f)$ to $U(2f)$, while  
the subgroup which leaves the mass term 
$<\bar q q>^2 + <q q >^2$ invariant is
O(2f) \cite{Kluberg}. 
By applying 
a generic O(2f) transformation to the  mass term
$\bar \chi \chi$ we identify the following set of operators
which might form condensates:

\begin{eqnarray}
\mbox{\rm scalar}\qquad\chi_1 \bar \chi_1 + \chi_2 \bar \chi_2 &\hskip 1 truecm 
&  
\mbox{\rm pseudoscalar}\qquad
\varepsilon(\chi_1 \bar \chi_1 + \chi_2 \bar \chi_2) 
\\
\mbox{\rm scalar diquark }\qquad\chi_1 \chi_2 - \chi_2 \chi_1 & \hskip 1 truecm 
& 
\mbox{\rm pseudoscalar diquark}\qquad
\varepsilon(\chi_1 \chi_2 -  \chi_2 \chi_1) \\
\mbox{\rm scalar antidiquark}\qquad\bar \chi_1 \bar \chi_2 - \bar \chi_2 \bar 
\chi_1  & \hskip 1 truecm&
\mbox{\rm pseudoscalar antidiquark}\qquad
\varepsilon (\bar \chi_1 \bar \chi_2 -\bar 
\chi_2 \bar \chi_1)
\end{eqnarray}
where the lower index labels colour.  The first line displays
the usual pseudoscalar and scalar condensates. The second (third) line
corresponds to diquark (antiquark) condensates, scalar and pseudoscalar. 
This simple minded quantum number assignment is done by considering
that  quark -- quark and quark-antiquark pairs have
opposite relative parity, and will be confirmed by the rigorous 
analysis below. See also the interesting discussions in \cite{SSV}.

Consider now quark propagation from a source at 0 to the point $x$.
The propagator $G_{ij}$ ($i,j$ color index) is an SU(2) matrix:

\begin{equation}
G_{ij} = \left(\matrix {a & b \cr
                   -b^\star & a^\star} \right)
\end{equation}
 
By taking its correlations in the appropriate
sector of quantum numbers we form meson and diquark 
operators. We shall limit ourselves to the local sector
of the spectrum and focus on the zero momentum {\it connected} propagators
of the scalar and
psudoscalar mesons and diquarks. The scalar meson propagator
will thus be an isovector, which we will call $\delta$, following QCD
notation.
 
The meson ($q\bar q$) and diquark ($qq$) and antidiquark ($\bar{q}\bar{q}$) 
propagators at 
$\mu=0$ are constructed from $G_{ij}$
as follows:-

\begin{eqnarray}
\mbox{\rm pion}\qquad \mbox{tr}GG^{\dagger}&=&  (a^2 + b^2) \label{eq:Gpion}\\
\mbox{\rm scalar meson}\qquad \varepsilon\mbox{tr} GG^{\dagger}
& =& \varepsilon (a^2 + b^2)\\
\mbox{\rm scalar $qq$}\qquad \mbox{det} G& =&  (a^2 + b^2) 
\label{eq:scalarqq}\\
\mbox{\rm scalar $\bar q\bar q$}\qquad  \mbox{det} G^{\dagger}& =& 
 (a^2 + b^2) \\
\mbox{\rm pseudoscalar $qq$}\qquad \varepsilon \mbox{det} G& =&
 \varepsilon (a^2 + b^2) \\
\mbox{\rm pseudoscalar $\bar q\bar q$}\qquad \varepsilon \mbox{det} 
G^\dagger
& =& \varepsilon (a^2 + b^2)
\label{eq:pseudoqq}
\end{eqnarray}

The notable feature of the propagators at $\mu=0$ is the exact degeneracies of
the pion, scalar $qq$ and scalar $\bar{q}\bar{q}$ and of the scalar meson,
pseudoscalar $qq$ and pseudoscalar $\bar{q}\bar{q}$. This can also
be understood as being due to quarks and anti-quarks having 
opposite intrinsic parities, whereas the $qq$ and $q\bar q$ interactions
due to gluon exchange are identical.
We then identify two orthogonal directions in the chiral space:\\
a) $\pi$ -scalar diquark - scalar antidiquark \\
b) $\delta$ - pseudoscalar diquark - pseudoscalar antidiquark

%




\subsubsection{$\mu \ne 0$}
When $\mu \ne 0$ the symmetry is reduced to that of staggered fermions, i.e.
$U_A(1) \times U_V(1)$ \cite{DMW}
, however the possibility of  diquark condensation
\footnote{ The conditions leading to the Vafa Witten theorem do not hold here.} 
could
lead to unusual patterns of chiral symmetry.

The condensate will
tend to rotate in chiral space as $\mu$ increases, 
rotating into a purely diquark direction for large $\mu$.  
However, as $\mu$ increases and the symmetry in chiral space is reduced,
as will be discussed below, 
the rotation will no longer be ``trivial'' (as it would be
at zero chemical potential) and the new vacuum would be physically
distinct from the original one. See also the interesting comments in
\cite{CD}.

Consider the Dirac kinetic operator $D{\!\!\!\! /}\,$ for staggered fermions. 
At $\mu=0$ the 
relation $D{\!\!\!\! /}\,^{\dagger}=-D{\!\!\!\! /}\,$ holds.
For $\mu\neq 0$  the fact that $e^{\mu}$ ($e^{-\mu}$) multiplies the 
forward (backward) timelike gauge
links in the lattice action means that ${D\!\!\!\! /}\,^{\dagger}
\neq-D{\!\!\!\! /}\,$, because in the matrix
$D{\!\!\!\! /}\,^{\dagger}$, 
$e^{-\mu}$ ($e^{\mu}$) now multiplies the forward (backward)
links. Consequently the propagators $G$ and $G^{\dagger}$
are no longer trivially related as they are at $\mu=0$. The composite 
propagators can be computed as follows\cite {KLS0}:
\begin{eqnarray}
\mbox{pion}\qquad & & \mbox{tr}G(\mu)G^{\dagger}(-\mu)\nonumber \\ 
\mbox{scalar $qq$}\qquad & & \mbox{det} G(\mu) \label{eq:nzdG}
\end{eqnarray}
with analogous formulas for the scalar meson and pseudoscalar $qq$.

Note for $\mu\not=0$, that although the diquark Goldstone modes in
(\ref{eq:dqGold}) contain both $\chi\chi$ and $\bar\chi\bar\chi$ operators,
we expect the physical states in the high density ground state to be
formed predominantly of quarks rather than anti-quarks.
We see from (\ref{eq:nzdG}) that 
for $\mu\neq 0$ the pion is no longer degenerate with the scalar $qq$; 
correspondingly the degeneracy of the scalar meson with the
pseudoscalar $qq$ is also broken.

Finally, consider the  ``baryonic'' pion  \cite{KLS} which is believed 
to be responsible for the failure of quenched 3 colours QCD (or indeed
for any treatment with a real path integral measure
proportional to $\mbox{det}M^\dagger M$). The propagator
would be:
\begin{equation}
\mbox{baryonic pion}\qquad \mbox{tr}G(\mu)G^{\dagger}(\mu) 
\end{equation}
We see that for the 2 colours theory a consequence of the fact that 
the quark propagators are SU(2) matrices even for $\mu\neq0$ is that 
the baryonic pion is degenerate with the scalar diquark via eqns.
(\ref{eq:Gpion},\ref{eq:scalarqq}): in SU(2) at nonzero chemical
potential it is still possible to express a diquark propagator as a positive
definite quantity.

\subsection{Rigorous discussion of symmetries and spectrum}

At zero density ($\mu=0$), we expect the model to display spontaneous 
chiral symmetry breaking, with a chiral condensate of the same form as the 
mass term
$m\sum_x\bar\chi(x)\chi(x)$. In terms of the $X,\bar X$ fields, introduced
in (\ref{eq:Xfields}),
\begin{equation}
\bar\chi\chi={1\over2}\left[\bar X_e\left(\matrix{&1\cr1&\cr}\right)
\tau_2\bar X_e^{tr}+
X_o^{tr}\left(\matrix{&1\cr1&\cr}\right)\tau_2X_o\right].
\label{eq:chcond}
\end{equation}
Of course, the condensate breaks the global symmetry (\ref{eq:U2}): the
residual symmetry is generated by the subgroup which leaves invariant the 
symmetric form $\left(\matrix{0&1\cr1&0\cr}\right)$, whose
most general element is $\left(\matrix{e^{i\alpha}&\cr&e^{-i\alpha}\cr}
\right)$, which generates $\mbox{U(1)}_V$. We thus identify the pattern of
chiral symmetry breaking as U(2)$\to$U(1) for $\mu=0$.

For $\mu=0$ the  number of broken generators is 3 (since dim U(2)=4), 
so we expect the spectrum of the model to contain three massless Goldstone 
modes. 
These 
can be found by considering infinitesimal rotations of the condensate
(\ref{eq:chcond}) by
$V_\delta=\One+i\delta\lambda$, with $\lambda$ one of the U(2) generators
$\{\One,\tau_i\}$, and identifying the mode as the coefficient of $O(\delta)$. 
The results are
\begin{eqnarray}
\One & \Rightarrow & \bar\chi\varepsilon\chi \nonumber\\
\tau_1 & \Rightarrow &
\chi^{tr}\tau_2\chi-\bar\chi\tau_2\bar\chi^{tr}\nonumber\\
\tau_2 & \Rightarrow &
\chi^{tr}\tau_2\chi+\bar\chi\tau_2\bar\chi^{tr}\nonumber\\
\tau_3 & \Rightarrow & \label{eq:Goldqbarq}\One
\end{eqnarray}
As expected, the first of these modes is the familiar pion 
$\bar\chi\varepsilon\chi$,
which is a pseudoscalar since it is odd under the lattice definition of parity:
\begin{eqnarray}
x=(x_0,x_1,x_2,x_3) &\mapsto& x^\prime=(x_0,1-x_1,1-x_2,1-x_3) \nonumber \\
\chi(x)\mapsto(-1)^{x_1^\prime+x_3^\prime}\chi(x^\prime),
&;&
\bar\chi(x)\mapsto(-1)^{x_1^\prime+x_3^\prime}\bar\chi(x^\prime).
\label{eq:parity}
\end{eqnarray}
The second two, however, are gauge invariant 
linear combinations of scalar diquark $\chi\chi$
and $\bar\chi\bar\chi$ states. The $\One$ indicates that rotations
generated by $\tau_3$ leave the condensate invariant. These states
carry baryon number and although they would be true Goldstone modes only 
in the chiral limit and at $\mu=0$, by continuity
we expect them to be light states in 
the low density regime with masses vanishing as $\sqrt{m}$.
For $\mu\neq 0$ of the three states identified only the pion remains
a Goldstone mode. The pion is generated by $\One$ and is associated
with spontaneous breaking of $U(1)_A$. 
The $\mbox{U(1)}_V$ rotation generated by $\tau_3$, as discussed above,
leaves the condensate invariant.

The symmetry breaking described here is notable because it is 
distinct from the $\mbox{SU}(2N_f)\to\mbox{Sp}(2N_f)$ predicted for 
continuum fermions in a pseudoreal gauge representation such as
the {\bf2} of SU(2)\cite{Peskin}. This discrepancy between lattice
and continuum has been known for some time  \cite{Kluberg}\cite{HT}.
For a complete classifications of the possible patterns of chiral 
symmetry breaking see \cite{Verb}. For patterns of 
chiral symmetry breaking the roles of fundamental and adjoint
representations appear to be reversed on the lattice \cite{SH}.

At high density ($\mu \gg m_{\pi}/2$), let us postulate that a large Fermi 
surface will promote
the formation of a diquark condensate. Which condensate forms is a dynamical
question, since in principle many diquark wavefunctions can be written
down. It is instructive to consider a ``maximally attractive channel'' (MAC)
for the formation of the condensate.
We proceed by making the following assumptions about the MAC condensate:
\begin{enumerate}

\item[(a)] The condensate is gauge invariant.

\item[(b)] The condensate is invariant under lattice parity (\ref{eq:parity})
and exhibits no preferred spatial orientation.

\item[(c)] The MAC condensate is as local as possible in the $\chi,\bar\chi$ 
fields.

\item[(d)] The condensate wavefunction is anti-symmetric under exchange of
fields.
\end{enumerate}
Of these four assumptions, only (d) should be considered inviolate
\cite{ARW}\cite{RSSV}. For isospinor fermions, however, 
it is straightforward to write
down a local condensate $qq_{\bf2}$ which satisfies (a) -- (d):
\begin{equation}
qq_{\bf2}={1\over2}\left[\chi^{tr}(x)\tau_2\chi(x)+
\bar\chi(x)\tau_2\bar\chi^{tr}(x)\right].
\label{eq:qq2}
\end{equation}
The relative + sign between the two terms is not arbitrary; this can be seen by
considering a two-point function 
$\langle\chi\chi(x)\bar\chi\bar\chi(y)\rangle$,
which in a vacuum with a diquark condensate will cluster at large separations
into $\langle\chi\chi\rangle\langle\bar\chi\bar\chi\rangle$ \cite{HM}
\cite{Sch}. We know 
that the scalar diquark propagator is proportional to the
determinant of an SU(2) matrix (see eqs.(\ref{eq:scalarqq} -
\ref{eq:pseudoqq}) above), and hence positive definite; therefore the
$\chi\chi$ and $\bar\chi\bar\chi$ condensates must form with the same sign.

In terms of the $X,\bar X$ fields the condensate is written
\begin{equation}
qq_{\bf2}={1\over2}\left[\bar X_e\left(\matrix{1&\cr&-1}\right)\tau_2
\bar X_e^{tr}+
X_o^{tr}\left(\matrix{1&\cr&-1}\right)\tau_2X_o\right].
\label{eq:dq2}
\end{equation}
The fascinating feature of this SU(2) model is that the diquark condensate 
(\ref{eq:dq2}) can be
obtained from the chiral condensate (\ref{eq:chcond}) by an explicit global
U(2) rotation:
\begin{equation}
V={i\over{\surd{2}}}\left(\matrix{1&1\cr-1&1\cr}\right).
\end{equation}
Therefore the pattern of symmetry breaking for $m=\mu=0$ remains 
U(2) $\to$ U(1), but the unbroken symmetry is no longer $\mbox{U(1)}_V$.
The chiral condensate is equivalent to the
diquark condensate due to the symmetry transformation relating them.
There
is only one discernable condensate
whose orientation in the chiral sphere is selected by the explicit
breaking term.
In lattice simulations we select the chiral condensate $\langle\bar qq\rangle$
because the bare mass $m\not=0$. 
In principle by setting the bare mass to zero and 
including a diquark source term we would obtain a diquark condensate 
$\langle qq\rangle$ at $\mu=0$.

For non-zero density one might naively expect that as
$\mu$ increases with respect to $m$  the condensate gradually {\sl rotates\/}
from a 
chiral $\langle\bar qq\rangle$ to a diquark $\langle qq\rangle$. 
Actually the true behaviour is more subtle, because
once $m\not=0$, $\mu\not=0$ 
the rotation is no longer 
trivial; the vacuum at large $\mu$ must be 
distinct from that at $\mu=0$. This fact is reflected by 
different numbers
of Goldstone modes in high and low density regimes; we shall argue
that there is a phase separation between low and high density,
with the diquark condensate acting as order parameter.  

We can check this by analysing the Goldstone modes obtained by U(2) rotations
of the diquark condensate (\ref{eq:dq2}):
\begin{eqnarray}
\One & \Rightarrow & \chi^{tr}\tau_2\varepsilon\chi+
\bar\chi\tau_2\varepsilon\bar\chi^{tr}\nonumber\\
\tau_1 & \Rightarrow & \One\nonumber\\
\tau_2 & \Rightarrow & \bar\chi\chi\nonumber\\
\tau_3 & \Rightarrow & \chi^{tr}\tau_2\chi-\bar\chi\tau_2\bar\chi^{tr}
\label{eq:dqGold}
\end{eqnarray}
Just as before, in the chiral limit for $\mu=0$, 
there are three Goldstone modes, 
a scalar $q\bar q$ 
state, a scalar $qq$ state (which is orthogonal to the condensate),
and a pseudoscalar $qq$ state. The 
spectrum of massless states has identical quantum numbers to those 
of (\ref{eq:Goldqbarq}) in this limit. The pseudoscalar
$q\bar q$, i.e. the pion, is {\sl no longer\/} a Goldstone particle in the
high density vacuum.
For $\mu \neq 0$ we retain two of these Goldstone modes: firstly the 
pseudoscalar diquark generated by $\One$
and associated with spontaneous breaking of
$\mbox{U(1)}_A$ which is a pseudo-Goldstone for $m\neq0$;
secondly the scalar diquark generated by $\tau_3$
and associated with spontaneous breaking of $\mbox{U(1)}_V$. This second state 
remains an exact Goldstone mode even once
$m\neq 0$, since there are no terms in (\ref{eq:sfund}) which explicitly
break this symmetry. Therefore we expect a level ordering 
(light $\to$ heavy) in the high density regime of 
scalar diquark (massless), pseudoscalar diquark, scalar meson, pion.
We will see below in sections \ref{sec:susc} and
\ref{sec:spec} to what extent these predictions are borne out in the
simulations. 



Recall that away from the chiral limit, 
in the low density regime 
but for $\mu\neq 0$ we
have no exact Goldstones, but only three light states, the pion and two 
scalar diquarks.
The fact that the low and
high density regimes are characterised by different numbers of massless
modes indicates that they should be separated by a true phase transition at 
some
$\mu=\mu_c$. This is indicated in the proposed phase diagram of
figure \ref{fig:phase2}. The Goldstone count in the diagram is as follows:
zero in the region to the right of the solid line; one in the region
to the left of this line where $\langle qq\rangle\not=0$; two in the chiral
limit along the $\mu$-axis; finally three at the U(2) symmetric point
at the origin.   

Finally, we should discuss the effects of using the hybrid molecular
dynamics \cite{DuKo} (or hybrid
Monte Carlo \cite{DKPR}) algorithm to simulate the model.
The functional integral measure contains a factor $\mbox{det}^N M^\dagger M$,
with $M$ the fermion matrix for one flavor of staggered fermion. Since
$\mbox{det}M^\dagger=\mbox{det}M^*=\mbox{det}\tau_2M\tau_2=\mbox{det}M$, 
then we know the simulation describes $2N$
identical flavors of staggered fermion. It can be shown that
 $\mbox{det}M$ is positive definite for $\mu\not=0$ \cite{Manfred}
therefore one could in principle ``take the square root'' by setting 
$N=\frac{1}{2}$ in  a hybrid algorithm.
To reduce the number of flavors 
in a simulation at zero chemical potential using staggered fermions, 
standard practice is to use even-odd 
partitioning. This relies on the equality of the fermion determinant
on the even sublattice with that on the odd sublattice which means that we 
require $\det M^{\dagger}=\det M$, 
a relation which no longer holds for $\mu\neq 
0$. 
Thus we are constrained to a minimum of $N_f=8$ 
continuum flavours in our simulation (recall that the number of physical
flavors $N_f=4N$ for staggered lattice fermions).
This should still result in an asymptotically-free theory,
since $N_f<{11\over2}N_c$; however, the two-loop coefficient of the
beta-function is now positive since $N_f>34N_c^3/(13N_c^2-3)$, which means
that the model should exhibit non-trivial fixed-point behaviour in the
far infra-red \cite{BZ}. The question of confinement/deconfinement is
thus further complicated. A recent non-perturbative
calculation by Appelquist and Sannino \cite {AS}
suggests indeed that the onset of the conformal phase is at
$N_f > 3.9 N_c$.
That would place our 8 flavours 2 colours results 
in the conformal phase; however, on the lattice volumes considered here
we have seen no 
indication of an 'exotic' behaviour.

The symmetries of the lattice model for $N=2$ flavours of staggered 
fermion will be very similar to the $N=1$ case discussed above 
because (i) the mass term and hence the chiral condensate will
be flavor symmetric, and (ii) the diquark channel we have analysed on
the assumption that it is MAC  must also be
flavor symmetric due to the Pauli exclusion principle. Therefore the whole
analysis should go through virtually unchanged, with the resulting pattern
of symmetry breaking for $\mu=0$ of U(4) $\to$ O(4), with 10 associated 
Goldstone modes.
Of course, all our measurement algorithms to date work in the single flavor 
sector, therefore we should only ``observe'' three Goldstone modes. 
So the special features of the SU(2) gauge theory arise from the fact that
the quarks and the anti-quarks belong to equivalent representations
which enlarges the continuous global symmetry 
of the lattice action describing $N$ flavors of
staggered fermion from $\mbox{U}(N) \times \mbox{U}(N)$ to $\mbox{U}(2N)$.  
The subgroup which leaves the mass term 
invariant is $\mbox{O}(2N)$.
In general there are  $4N^2 -N(2N-1) = 2N^2 + N$ broken generators
associated with $N^2$ Goldstone mesons, and $N^2 + N = 2N(N+1)/2$ Goldstone
baryons and antibaryons.
It is instructive to consider what will happen as the flavor symmetries enlarge
in the continuum limit,
when we expect the model to describe $N_f=4N$ quarks. It is currently unclear 
whether the lattice model in this limit will persist
in the non-standard $\mbox{SU}(2N_f)\to\mbox{O}(2N_f)$ (note
the axial anomaly should reemerge in this limit), or assume the 
orthodox continuum pattern $\mbox{SU}(2N_f)\to\mbox{Sp}(2N_f)$ \cite{Peskin}.

\section{Simulations}

In this initial study of two colours QCD at non-zero chemical potential
we used a standard hybrid molecular dynamics algorithm \cite{DuKo,muzerohyb}. 
Since we cannot
use even-odd partitioning when $\mu\neq 0$ we were constrained to the minimal
value of 8 continuum flavours.

  We simulated at $\mu\neq 0$ at
couplings: $\beta=1.3$ where the system was in the phase of broken
chiral symmetry at $\mu=0$. Most of 
the simulations were on $6^4$ lattices however we studied the
 mass spectrum on a $6^{3}\times12$ lattice at $\beta=1.3$ with $m=0.07$ for 
$\mu=0.0, 0.2, 0.4$. Note that although the fermion determinant of the SU(2)
theory is positive definite for $\mu\neq 0$, it is likely that value of the
fermion determinant has large fluctuations in magnitude when the chemical 
potential
is switched on. In support of this conjecture we found that the simulations 
numerically intensive for $\mu>m_\pi/2$.

For chemical potential smaller than $m_\pi/2$ simulations were not problematic
and in this case we chose dt = .02 and ran for 5000
sweeps, after discarding 1000 sweeps for thermalization.

For $\mu\ge m_{\pi}/2$, we observed several instabilities.
 Typically, the system
was crashed after a few tens of iterations. To combat this we  refreshed the 
system frequently (every 5 sweeps for the gauge fields and 3 for the fermions).
In this way we could run for thousands of iterations with dt = .02. With
the exception of $\mu = .8$ , dt = .05 worked also. 
As final strategy, we thermalized with dt=.05 for 500 iterations, subsequently
reduced dt to .02 , reequilibrated for 300 sweeps proceeded to measure for 5000 
sweeps.

To monitor any possible bias introduced by the frequent refreshment 
we also continued our runs at $\mu = .6$ with a pure molecular dynamics
evolution, which revealed no problem for a few thousands steps 
(therefore we stopped it).

Clearly, we cannot exclude ergodicity problems.

\begin{figure}
\begin{center}
{\epsfig{file=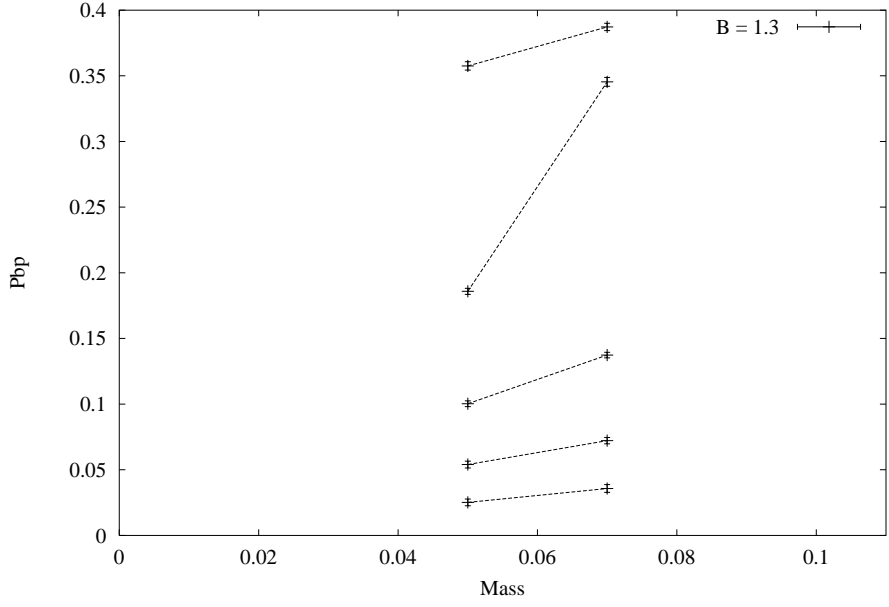, width=12 truecm}}\\
\end{center}
\caption[xxx]{$\langle\bar qq\rangle$ as a function of the bare mass
for $\beta = 1.3$ (upper)  
at $\mu = 0, .2, .4, .6, .8$} \label{fig:pbp_vs_mass}
\end{figure}

In Figs. \ref {fig:pbp_vs_mass}
we show the chiral condensate as
a function of bare mass $m$ for a range of values of chemical
potential. The system
is clearly in the broken phase at $\mu=0.0$ and the chiral condensate 
decreases quite sharply as $\mu$ is increased. 

In the standard scenario, 
we expect that as $\mu$ is increased beyond $m_{\pi}/2$ the 
ground state populated by quarks will be favoured over the vacuum 
state; therefore chiral symmetry may be restored for $\mu\ge m_{\pi}/2$. 
We found that at $\beta = 1.5$ for $m = 0.1$, the pion mass $m_{\pi}\simeq 0.8$
while at  $\beta = 1.3$ with $m = 0.07$ we find $m_{\pi}\simeq 0.6$

\begin{figure}
\begin{center}
{\epsfig{file=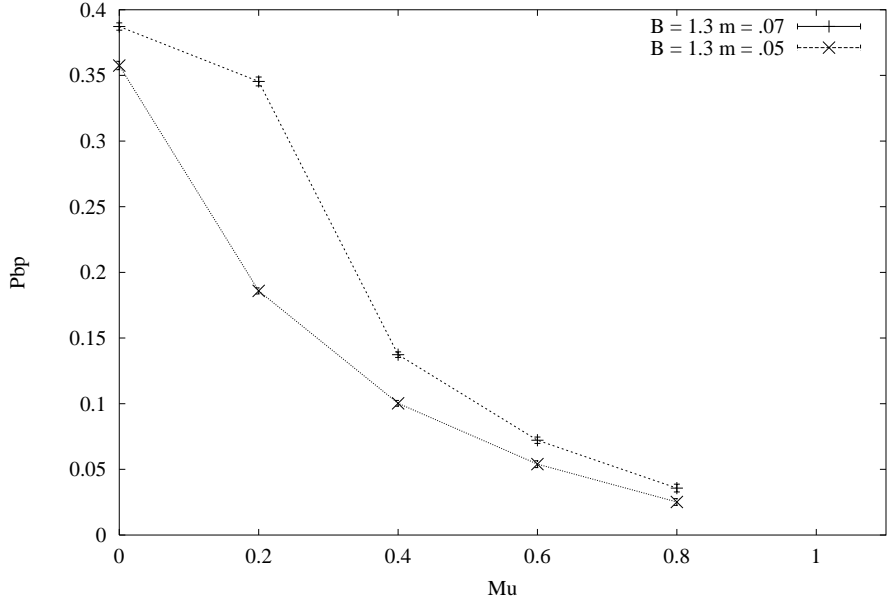, width= 12 truecm}}
\end{center}
\caption[xxx]{$\langle\bar qq\rangle$ as a function of the 
chemical potential for $\beta = 1.3$
and masses as shown in the diagrams}
\label{fig:pbp_vs_mu}
\end{figure}

Behaviour consistent with the standard scenario is reflected in the 
Figs. \ref {fig:pbp_vs_mu} where the chiral condensate is 
plotted as a function 
of $m$. At $\beta=1.3$ and at $\mu=0.2$ there is a strong 
$m$-dependence in the value of the chiral condensate. The chiral 
transition seems significantly sharper at $m=0.07$ than it is at $m=0.05$
as we would expect.  See the spectrum section \ref{sec:spec} for a discussion
of these results.

\section{Particle susceptibilities}
\label{sec:susc}

Measurement of the meson and diquark susceptibilities will
allow us to assess the effects of chiral symmetry restoration 
and, to some extent, explore the level ordering in the particle spectrum
\cite{KKL}
which should reflect properties noted in the earlier  discussion 
of symmetries where we identified Goldstone modes in high and low density
phases.

In terms of the timeslice propagator $P(t)=\sum_{\bf x}G^{'}({\bf x},t)$,
the susceptibility, $\chi$, is defined by 
\begin{equation}
\chi=\sum_{t}\;P(t).
\end{equation}
On the assumption that a particular channel is dominated by a single pole, 
then $\chi$ can be written $Z/M_{bs}^2$, where $M_{bs}$ is a bound state mass
and $Z$ is a wavefunction renormalisation constant. If we make a second
assumption that the $Z$'s are relatively insensitive to the quantum number of
the state, then susceptibility measurements yield information about level
ordering without the need for any propagator fitting, and may be a more 
reliable guide on the  small lattice volumes presented here (see however
the comments below).

 We have measured susceptibilities of the scalar mesons and diquarks,
taking into account contributions from connected diagrams only,
and plotted the quantity $m\chi$ 
which we expect (from the Ward
identity) to be equal 
to the chiral condensate in the case of the pion susceptibility.
Although the realization of PCAC in this model has yet to be
worked out in detail, it is reasonable to assume that
in each channel the susceptibility gives a direct measure of
the associated condensate.

The degeneracies of the meson and diquark sectors are broken for any
$\mu\neq0$.
The pion which is a pseudo-Goldstone for $\mu<m_{\pi}/2$, becomes heavy
in the dense phase whereas the pseudoscalar $qq$ state which is heavy
in the low density phase, becomes light in the high density phase. 
Recall
from section \ref{sec:symmetries} that we expected the scalar diquark to be an
exactly massless mode in the high density phase. 
Unfortunately it is impossible to infer the level ordering directly from the
susceptibilities, since the possible occurrence of a plateau prevents it (see
the discussion in Sec. \ref{sec:spec}).



\begin{figure}
\begin{center}
{\epsfig{file=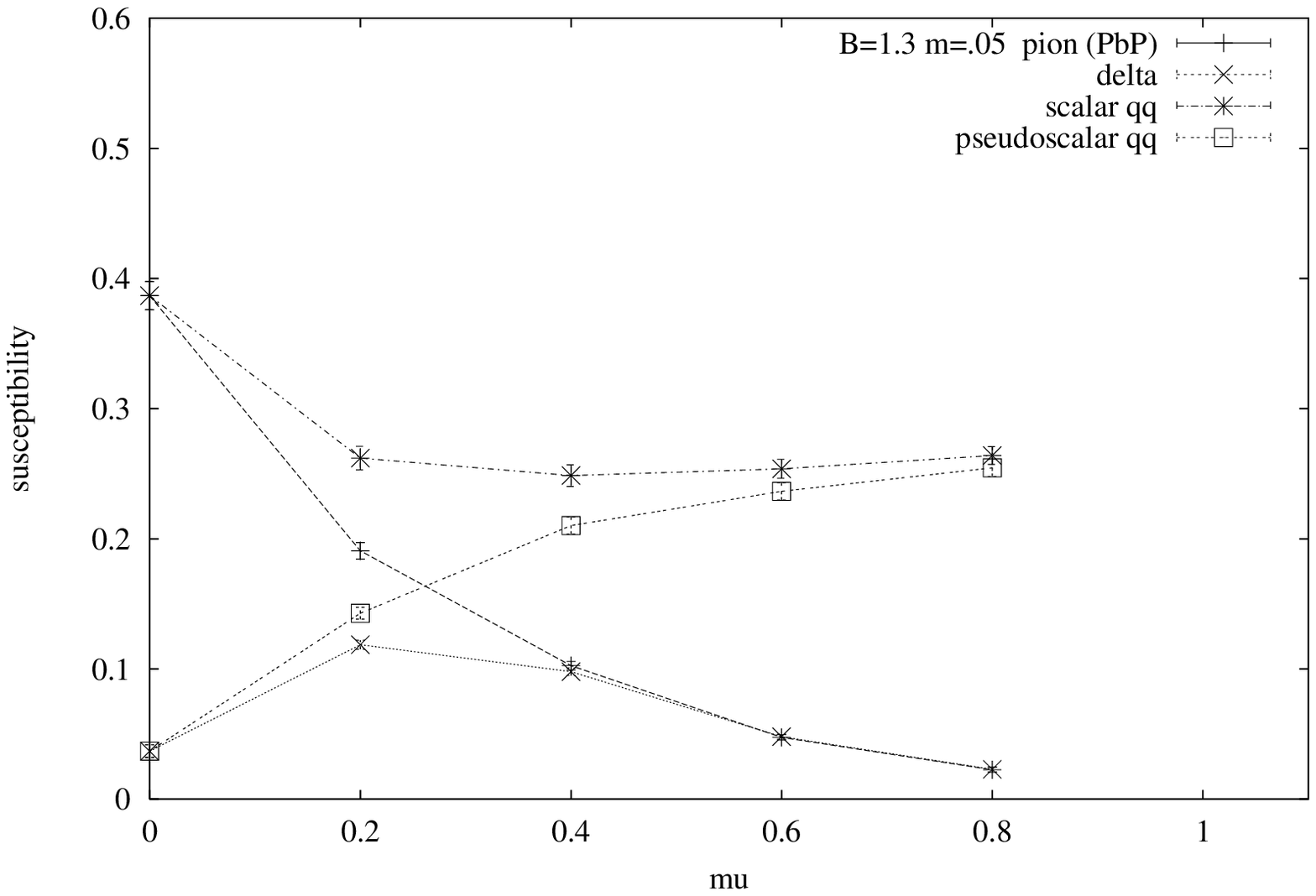, width = 12 truecm}
\epsfig{file=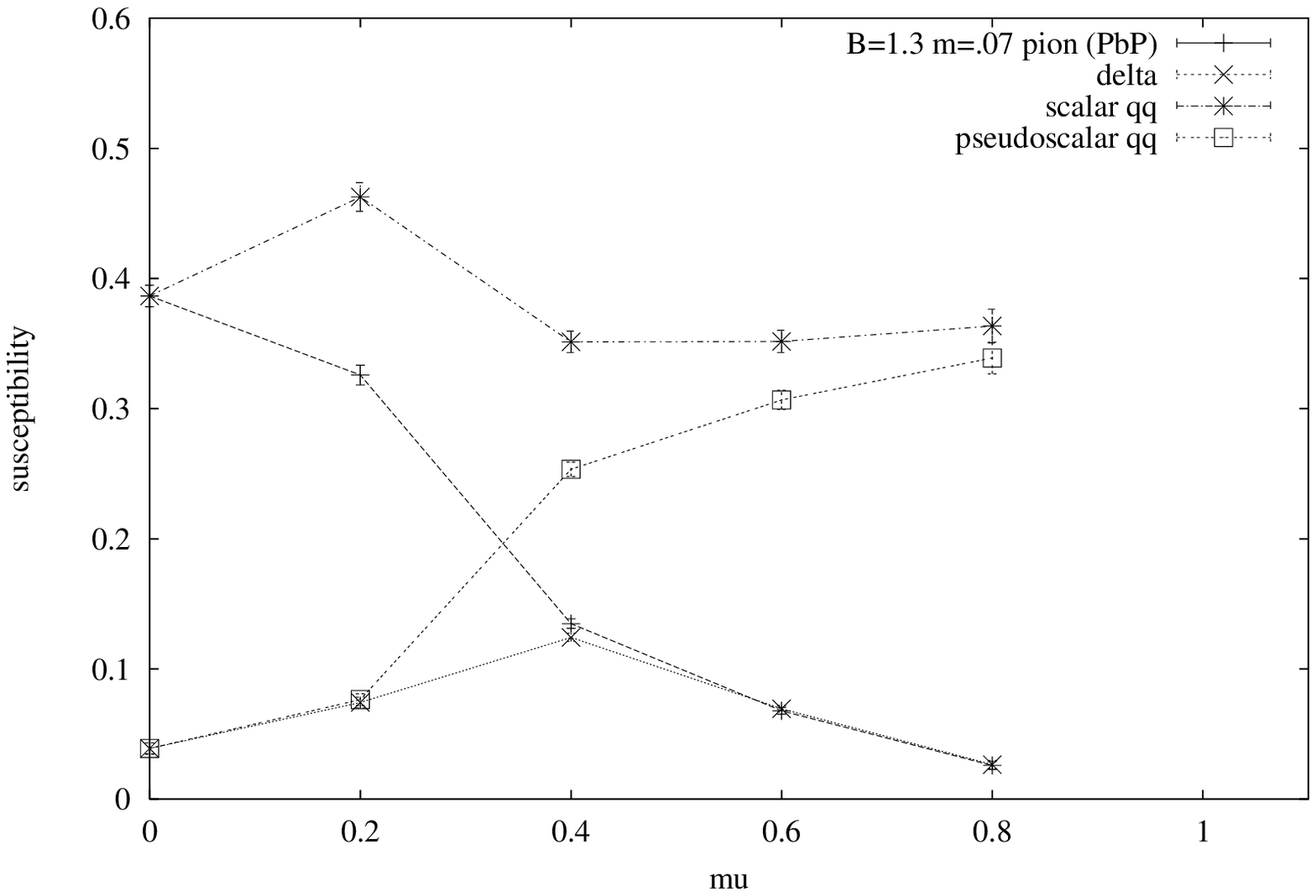 , width = 12 truecm}}
    \caption[xxx]{Scalar and pseudoscalar susceptibilities 
for mesons and diquarks
a function of $\mu$ at $\beta = 1.3$ and mass = .05
(upper) and mass = .07. See text for discussions}
    \label{fig:mu_I_b1.3}
\end{center}
\end{figure}

The level ordering
we infer (lightest to heaviest) from Fig. \ref{fig:mu_I_b1.3}
for the susceptibilities 
is scalar diquark, pseudoscalar diquark,
and a degenerate pion and delta susceptibilities. 

Degeneracy of the pion and $\delta$ is usually interpreted 
as a signal for the restoration of $U_A(1)$ symmetry. It would be natural
to infer from that a complete suppression of the topological fluctuations
induced by the ordering effects of the chemical potential \cite{ua1}.

Rigorous arguments
predict that, indeed, the scalar diquark propagator should always be
the larger one \cite{SZ}. However, it is not possible to infer from
this a mass level ordering since the scalar propagator has a disconnected
component, and because of the multiplicative renormalization of
the wavefunction. 
For these reasons we have performed a mass spectrum calculation
which will show that the particle content of the scalar and pseudoscalar 
diquark propagators is indeed the same and we  postpone a detailed 
comparison of the numerical and analytic results to the spectrum section.

\section{Particle Spectrum}
\label{sec:spec}

To extend and verify the information obtained from the susceptibilities
we have performed particle spectrum measurements on a lattice with
extended temporal direction. We simulated at $\beta = 1.3$, $m=0.07$
on a $6^3 \times 12$ lattice. The qualitative trends revealed in the 
susceptibility measurements are the fact that we have a light pion 
and scalar diquark in the low density phase and two light diquark
states in the high density phase. We expect the pion to become heavy 
and degenerate with the connected contribution to the scalar meson
(called $\delta$) for  $\mu>m_{\pi}/2$.

Finite density spectroscopy analysis 
has been introduced and discussed
in \cite{KLS0}. The following symmetries should hold
true in the ensemble:
\begin{equation}
G(t) = G(T-t) \qquad \mbox{\rm for mesons}
\end{equation}
\begin{equation}
G(t,\mu) = G(T-t, -\mu) \qquad\mbox{\rm for diquarks}
\end{equation}

The particle propagators for $\mu=0.0$ and for $\mu=0.4$ are shown in
Fig. 6.
At $\mu=0$  the pixon and scalar diquark are degenerate
light states while the pseudoscalar diquark and scalar meson 
are degenerate but comparatively heavy. The propagators are
symmetric in time at $\mu=0$. However at $\mu=0.4$ 
the chemical potential induces an asymmetry in the 
propagators by favouring forwards over backwards propagation.
As expected from the susceptibilities, the pion is now
a heavy state which is approximately degenerate with the scalar meson.
We also note (see Figs. 6 and \ref{fig:diffe} ) that at $\mu = 0.4$
$G_{\rm scalar\;qq}(t) = G_{\rm pseudoscalar\;qq}(t) + p$, where $p$ is 
a constant. This means that the pole contents of scalar and pseudoscalar
diquark propagators are identical, i.e. 
the scalar and pseudoscalar diquarks are
degenerate. We found that $p$ increased  as $\mu$ was
increased, this trend being more clear in the scalar channel
where condensation is expected
(Fig. \ref{fig:pla}). If this trend were to persist for a range
of lattice volumes then it  could be intepreted as evidence
for condensation in the scalar diquark channel with
$ p \propto \vert\langle qq\rangle\vert^2$. 
A similar analysis has been applied to simulations of the 2+1 dimensional 
Gross-Neveu model \cite{HM}; 
stable plateaux were indeed observed, but the
volume scaling of the constant $p$ did not show the expected behaviour.
To confirm this scenario one would have to 
implement an explicit diquark source term $jqq$ and extrapolate the source 
$j$ to zero: this work is in progress \cite{MH}.

The dashed lines in Figs. 6 
are the results of the fits to
\begin{equation}
G_m(t) = a(e^{(-m t)} + e^{m(T-t)}) + (-1)^tb(e^{(-m_1 t)} + e^{m_1(T-t)})
+ p 
\end{equation}
for the mesons and
\begin{equation}
G_b(t) = ae^{-m_{f} t} + be^{m_{b}(T-t)} + p 
\end{equation}
for the diquarks.

For the mesons $p$ should be zero and $m$ and $m_1$ correspond to masses
of states with different lattice parity. Since the mesons do
not ``feel'' the effects of the chemical potential the masses of the forward 
and backward propagating meson states are identical.
Although the inclusion of the oscillating term was
mandatory for a good fit, 
very often only the fundamental amplitude and the first mass were
meaningfully determined.
For $\mu\neq 0$ we expect $m_{f} \ne m_{b}$ in the diquark channels,
reflecting the different backwards and forward propagations in the 
dense medium. 

The particle masses for the mesons and for the diquark states 
were measured at $\mu=0.0,0.2,0.4$ and are plotted in 
Fig. \ref{fig:mass}. The pion mass increases, while the
scalar meson mass decreases as $\mu$ is increased and the two
particles become approximately degenerate at $\mu\simeq 0.4$
(at zero density $m_{\pi}/2 \simeq 0.3$). The scalar diquark which
is degenerate with the pion at $\mu=0.0$ is already lighter than 
the pion at $\mu=0.2$ and by $\mu=0.4$ the  pseudoscalar and scalar 
diquarks are apparently considerably lighter than the pion and delta meson.

In conclusion the ordering is as follows: 
two light(see however the comment below), degenerate 
diquarks, two heavier, degenerate pion and delta 

This ordering is not quite what was predicted in section
\ref{sec:symmetries}, and several comments are in order.
First, recall that the predictions of section \ref{sec:symmetries}
are for the $\sigma$ meson, while the numerical results are
for the $\delta$ meson. It is certainly possible that the $\sigma$
is lighter.

Next, the operator (\ref{eq:scalarqq}), being manifestly
real, may not project onto the true Goldstone mode, which corresponds to 
fluctuations in the phase of the condensate. It is interesting to note that 
for diquark condensation, unlike chiral condensation, there is no spacetime
quantum number (such as parity) which serves to distinguish transverse (ie.
Goldstone) from longitudinal (ie. Higgs) excitations.

A speculative comment is in order.
A naive approach to calculating masses in the presence of a chemical
potential would suggest that the zero density masses would be shifted 
downwards by 
$n \times \mu$,
where $n$ is the number of quarks in the state under consideration. 
This shift merely reflects a change of the
reference level of the energy, therefore does not take into account
any new dynamics associated with propagation through a dense medium.
To appreciate the true dynamical content we plotted, together with
the masses themselves, the reference energy $E(\mu) = E(0) - 2\mu$.
We do indeed observe that the change of the masses is genuine and not merely
a simple shift in the energy scale.
However, if we add a shift $n\mu$ to our spectrum results,  
the data suggest that all the four masses are
then nearly degenerate. Note that preliminary calculations in a model
field theory suggest precisely this result \cite{SZ}.]

Finally, if we estimate the constituent 
mass from the $\rho$ propagator, as seems
reasonable, we would convince ourselves that these states are still bound 
as $\rho \simeq 1.8 > \pi \simeq 1.2$. 
In conclusion, the results for the $\pi$--$\delta$, and the 
scalar--pseudoscalar diquarks, 
masses  suggest  degenerate, composite, massive, but rather light
particles. 

\section{Summary and outlook}

  In this  study of two colors QCD at non-zero
chemical potential with dynamical fermions we have outlined in
detail the symmetries of the lattice model and identified the
Goldstone modes in the low and high density phases incorporating the
possibility of diquark condensation for $\mu>m_{\pi}/2$.  
In our preliminary simulations we have established
the level ordering of the particle spectrum from the 
susceptibilities and via direct measurements for several values
of $\mu$. 
The analytic results predict Goldstone states which we have not
observed in the simulations. 
The simulation results indicate four degenerate states at
large $\mu$, when the data are referred to the same energy scale, or
two couples of degenerate particles if we do not change the energy
reference.  As discussed in the text, it is conceivable that none
of the operators we have implemented are suitable for measuring
the Goldstone modes predicted by the analytic calculations.
We have noted exact
degeneracies at $\mu=0$ between pion and scalar diquark; scalar
meson and pseudoscalar diquark. The pion which is a pseudo-Goldstone
in the low density phase becomes heavy in the dense phase whereas
the pseudoscalar diquark is heavy in the low density phase but
becomes light in the high density phase. For $\mu>m_{\pi}/2$ the
pion and scalar meson $\delta$ are approximately degenerate.
It would be very interesting to measure the $\sigma$ which would
require the disconnected contributions. It is certainly possible
that the sigma and delta masses are different.  

 It is possible that the rotation of chiral
condensate to baryonic condensate first noted in the infinite
coupling regime \cite{KarDag} persists in the intermediate and
weak coupling regimes and our symmetries analysis supports this.
The fact that the scalar and pseudoscalar diquark states are found
to be light for $\mu>m_{\pi}/2$ is also suggestive.
 In an extension of this work we will implement an explicit diquark
source term in a hybrid Monte Carlo simulation and investigate the 
effects of extrapolating the source to zero \cite{MH}.  

Finally, we notice that while the diquark  observables introduced
for studying condensation phenomenon 
display an interesting and peculiar
behaviour, there is nothing in the behaviour of the
conventional observables discussed here 
- - chiral condensate, meson spectrum - which
requires going beyond the standard scenario. The thermodynamics of
this model is considered in a companion paper \cite{PH}.

\section*{Acknowledgements}
MPL, SEM and SJH received support from EU TMR contract no. ERBFMRXCT97-0122.
SEM thanks the Zentrum f\"ur Interdisziplin\"are Forschung der Universit\"at
Bielefeld for hospitality during the early stages of this project. JBK thanks
the National Science Foundation ( NSF-PHY96-05199 ) for partial support.

\newpage

\newpage

\newpage
\begin{figure}
\begin{center}
{\epsfig{file=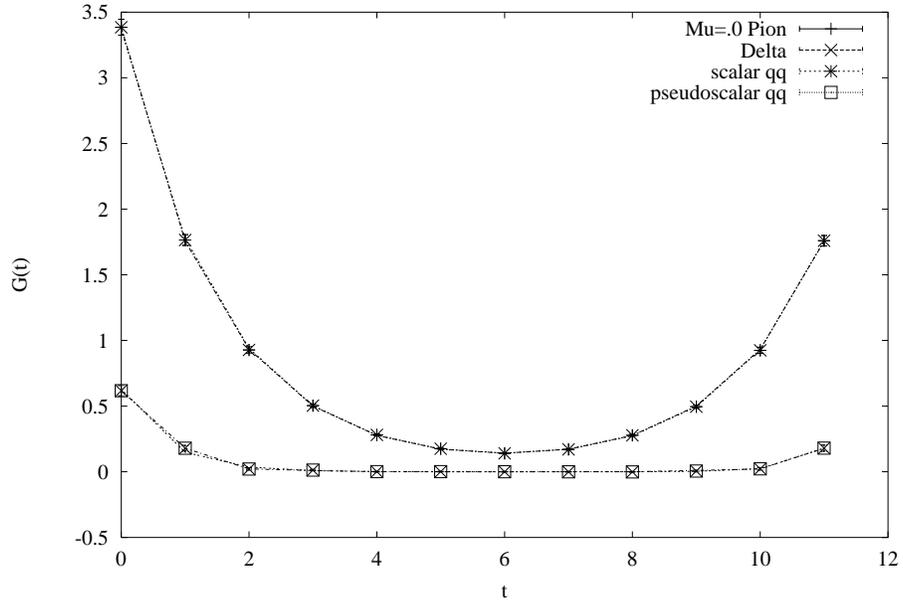, width = 12 truecm} \\
\epsfig{file=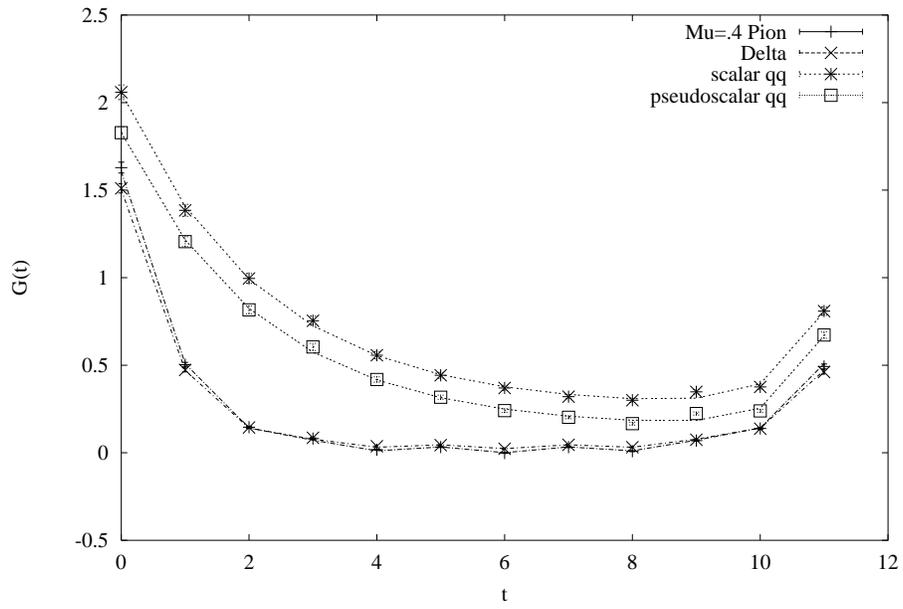, width = 12 truecm}}
\caption[xxx]{Level ordering at $\mu = 0.0$ and $\mu = 0.4$}
\end{center}
\label{fig:pss}
\end{figure}%

\newpage
\begin{figure}
\begin{center}
{\epsfig{file=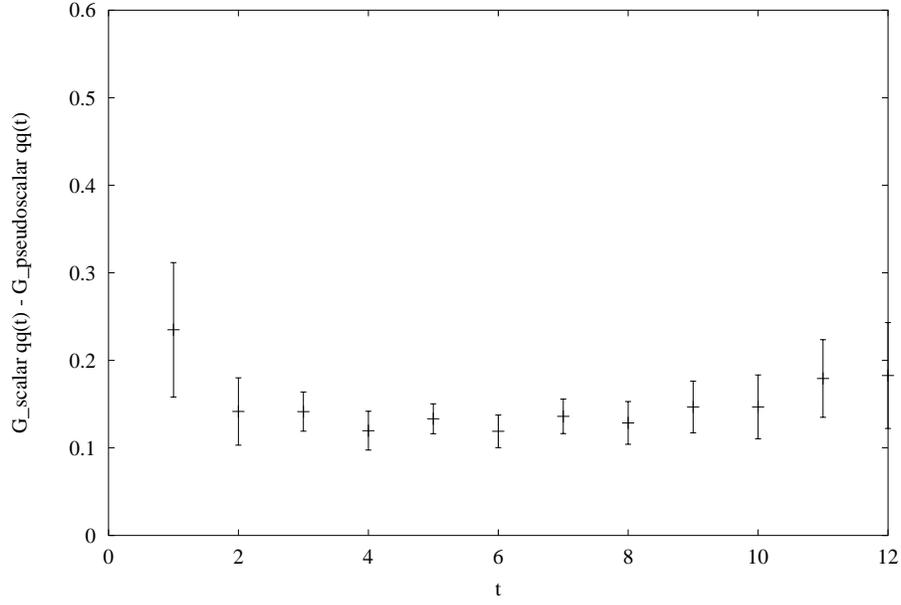, width = 12 truecm}}
\end{center}
\caption{Difference of the scalar and pseudoscalar diquark propagators
as a function of the time distance, suggesting a) condensation in
the scalar channel b) similar particle content in the two channels. }
\label{fig:diffe}
\end{figure}%

\begin{figure}
\begin{center}
\epsfig{file=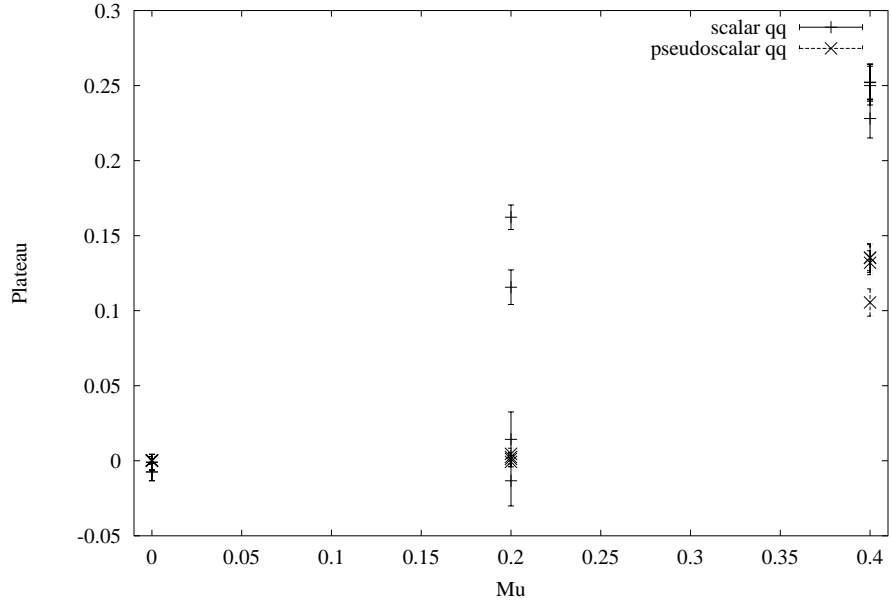, width = 12 truecm}
\end{center}
\caption{Plateau p ( $\simeq <\chi \chi> ^2$  )
from the fits of the scalar and pseudoscalar propagators}
\label{fig:pla}
\end{figure}

\newpage
\begin{figure}
\begin{center}
{\epsfig{file=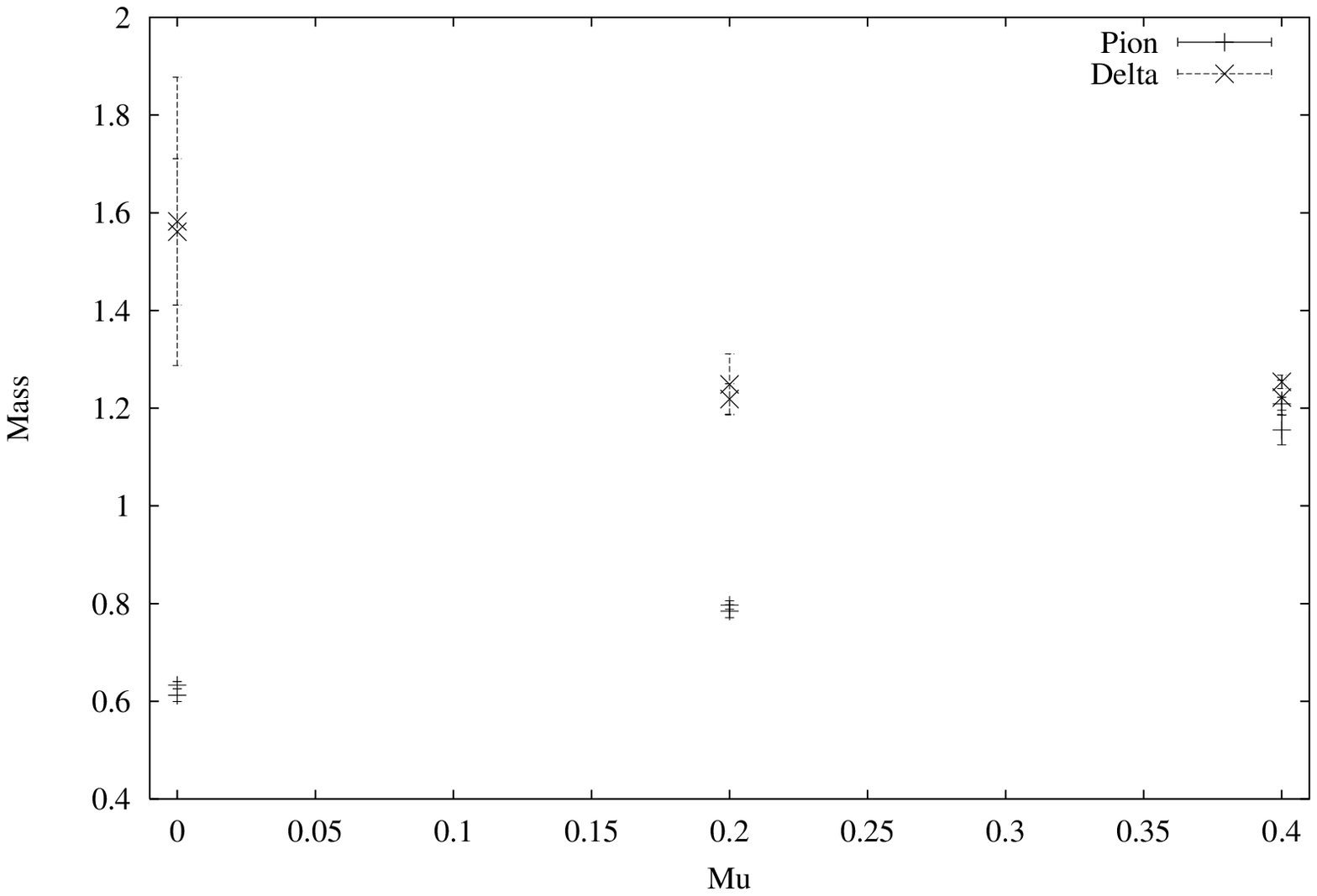, width = 12 truecm} \\
\epsfig{file=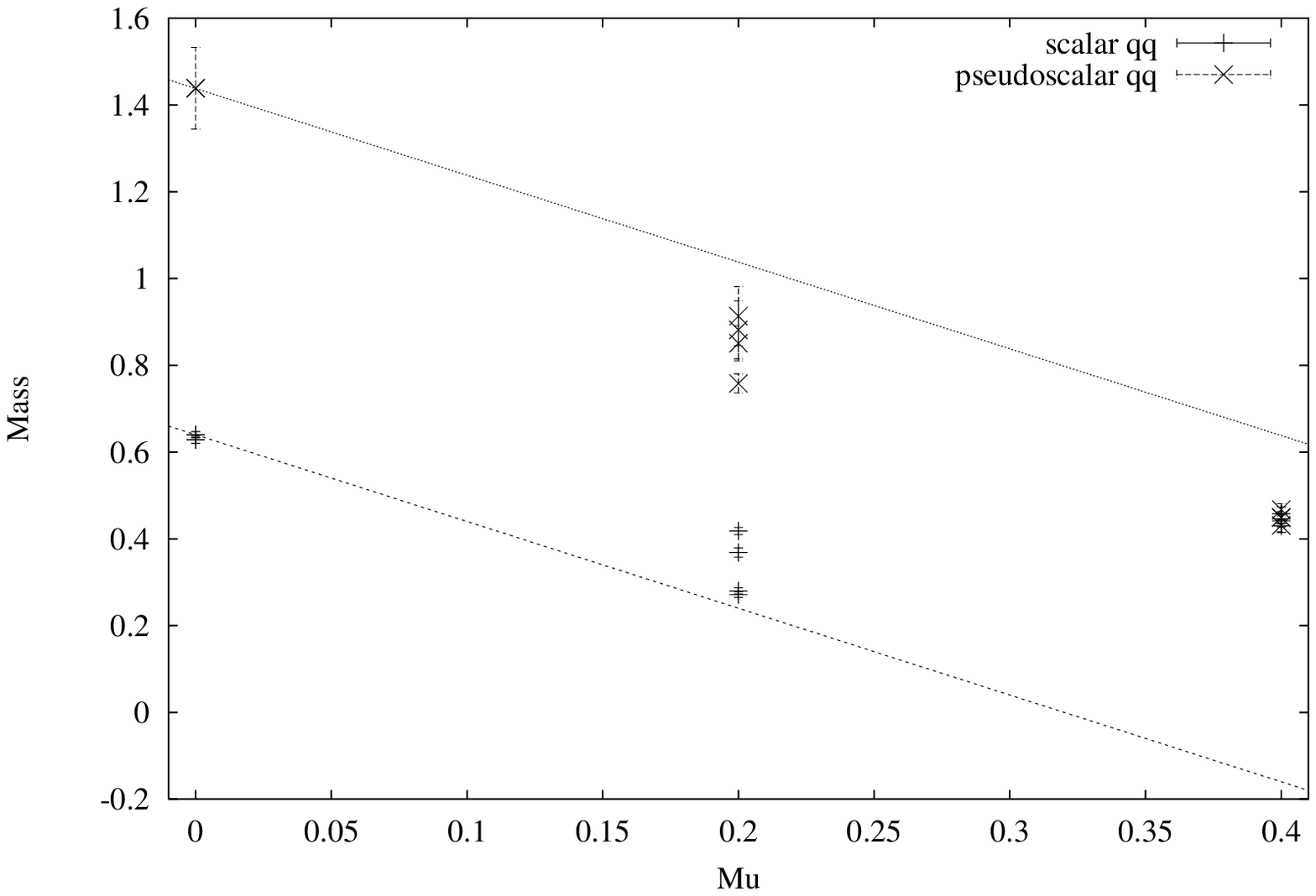, width = 12 truecm}}
\end{center}
\caption{Masses as a function of the chemical potential.
For the diquarks we also show the naive prediction
$m (\mu) = m(\mu=0) - 2*\mu$}
\label{fig:mass}
\end{figure}%


\begin{thebibliography}{99}

\bibitem{DuKo} S. Duane and J.B. Kogut, Nucl. Phys. {\bf B275} (1986) 398.
\bibitem{DKPR} S. Duane, A.D. Kennedy, B.J. Pendleton and D. Roweth,
Phys. Lett. {\bf B195} (1987) 216.
\bibitem{KL} See e.g.  Proceedings of {\it QCD at Finite Baryon Density},
Bielefeld, April 1998, F. Karsch and M.-P Lombardo, eds, Nucl. Phys. {\bf A
642}, 1998.
\bibitem{BarBel92}I.M. Barbour and A.J. Bell, Nucl. Phys. {\bf B372} 
(1992) 385.
\bibitem{ARW} M. Alford, K. Rajagopal and F. Wilczek, Phys. Lett. {\bf B422}
(1998) 247; Nucl. Phys. {\bf B537} (1999) 443. 
\bibitem{RSSV} R. Rapp, T. Sch\"afer, E.V. Shuryak and M. Velkovsky,
Phys. Rev. Lett. {\bf81} (1998) 53.
\bibitem{KarDag} E. Dagotto, F. Karsch and A. Moreo, Phys. Lett. 
{\bf 169B} (1986) 421.
\bibitem{DMW} E.Dagotto, A.Moreo, U.Wolff, Phys. Lett. {\bf 186B} (1987)
395.
\bibitem{quensu2} C. Baillie et.al., Phys. Lett. {\bf 197B} (1987) 195.
\bibitem{CD} G.W. Carter and D. Diakonov,  hep-ph/9812445 and references
therein. 
\bibitem{Kluberg} H. Kluberg-Stern, A. Morel and B. Petersson,
Nucl. Phys. {\bf B215} (1983) 527.
\bibitem {SSV} T. Sch\"afer, E.V. Shuryak and J.J.M. 
Verbaarschot,  Nucl.Phys. {\bf B412}
(1994) 143.
\bibitem{KLS} M.-P. Lombardo, J.B. Kogut and D.K. Sinclair, 
Phys.Rev. {\bf D54} (1996) 2303. 
\bibitem{KLS0} J.B. Kogut, M.-P. Lombardo and D.K. Sinclair, 
Phys. Rev. {\bf D51} (1995) 1282.
\bibitem{Peskin} M.E. Peskin, Nucl. Phys. {\bf B175} (1980) 197.
\bibitem{HT} S.J. Hands and M. Teper, Nucl. Phys. {\bf B347} (1990) 819.
\bibitem{Verb} J.J. M. Verbaarschot, Phys.Rev.Lett. {\bf 72} (1994);
M.A. Halasz and J.J.M. Verbaarschot, Phys. Rev. Lett.
{\bf 74} (1995) 3920.
\bibitem{SH} S. Hands, in preparation. 
\bibitem{HM} S.J. Hands and S.E. Morrison, hep-lat/9807033.
\bibitem {Sch} Th. Schaefer, Nucl.Phys.{\bf A642} (1998) 45.
\bibitem{Manfred} Private communication with M. Oevers and I. Barbour.
\bibitem{BZ} T. Banks and A. Zaks, Nucl. Phys. {\bf B196} (1982) 189.
\bibitem{AS} T. Appelquist and F. Sannino, hep-ph/9806409.
\bibitem{muzerohyb}J.B. Kogut, J. Polonyi, H.W. Wyld, D.K. Sinclair,
Phys. Rev. {\bf D31} (1985) 3307.
\bibitem{KKL} A. Koci\'{c}, J.B. Kogut and M.-P. Lombardo, Nucl. Phys.
{\bf B398} (1993) 376.
\bibitem{ua1} E.V. Shuryak, Comments Nucl. Phys. {\bf 21}, (1994) 235.
\bibitem{SZ} M. Stephanov and A. Zhitnitsky, work in progress, 
communicated by M. Stephanov.  
\bibitem{MH} S.E. Morrison and S.J. Hands, contribution to the workshop
{\sl Strong
and Electroweak Matter\/},
Copenhagen, 2nd - 5th December 1998, hep-lat/9902012.
\bibitem{PH} M.-P. Lombardo, to be submitted.

\end{thebibliography}
\end{document}